\documentclass[english,twoside,12pt]{article}
\usepackage{comment}
\usepackage{float}
\usepackage[latin1]{inputenc}
\usepackage[T1]{fontenc}
\usepackage{amsmath}
\usepackage{amsfonts}
\usepackage{graphicx}
\usepackage{subfigure}
\usepackage{a4wide}
\usepackage{amssymb}
\usepackage{fancyhdr}
\usepackage{mathrsfs}
\DeclareMathOperator{\arcsinh}{arcsinh}
\usepackage[toc,page]{appendix}

\newcommand{\be}{\begin{equation}}
\newcommand{\ee}{\end{equation}}

\newcommand{\bea}{\begin{eqnarray}}
\newcommand{\eea}{\end{eqnarray}}

\newcommand{\non}{\nonumber}

\begin{document}

\title{Non-Singular Black Holes and mass inflation in modified gravity} 

\author{ 
Manuel Bertipagani$^{1}$\footnote{E-mail address: manuel.bertipagani@studenti.unitn.it},\,\,\,
Massimiliano Rinaldi$^{1,2}$\footnote{E-mail address: massimiliano.rinaldi@unitn.it},\,\,\,\\\\
Lorenzo Sebastiani$^{3}$\footnote{E-mail address: lorenzo.sebastiani@pi.infn.it},\,\,\,
Sergio Zerbini$^{1}$\footnote{E-mail address: sergio.zerbini@unitn.it}
\\
\\
\begin{small}
$^1$Dipartimento di Fisica, Universit\`a di Trento, Via Sommarive 14, 38123 Povo (TN), Italy.
\end{small}\\
\begin{small}
$^2$TIFPA - INFN,  Via Sommarive 14, 38123 Povo (TN), Italy.
\end{small}\\
\begin{small}
$^3$Istituto Nazionale di Fisica Nucleare, Sezione di Pisa, Italy.
\end{small}
}
\date{}

\maketitle{}

\abstract{
We analytically derive a class of  non-singular, static and spherically symmetric topological black hole metrics in $F(R)$-gravity. These have not a de Sitter core at their centre, as most model in standard General Relativity. We study the geometric properties and the motion of test particles around these objects. Since they have two horizons, the inner being of Cauchy type, we focus on the problem of mass inflation and show that it occurs except when some extremal conditions are met.}

%----------------------------------------------------------------------------------------
%	INTRODUZIONE
%----------------------------------------------------------------------------------------
\section{Introduction}

The recent detection of gravitational waves from binary systems of black holes (BHs) and the ``multimessanger'' 
signals from the first observation of the collision of two
relativistic neutron  stars largely confirmed some predictions of the strong-field regime of General Relativity (GR) \cite{LIGO1,LIGO2, LIGO3, LIGO4}. Moreover, two years ago, the Event Horizon Telescope collaboration\cite{HT} published the first imagine of a supermassive black hole at the center of M87. It is then fair to say that the existence of black holes is almost certain, which makes longstanding conceptual problems, such as the central singularity and the information paradox, even more pressing.

Usually, astrophysical black holes are described by the Kerr metric, which is a stationary, vacuum solution of the field equations of GR, with a ring-shaped singularity at the center. Since space-time singularities are problematic, a lot of investigation has been dedicated to viable alternatives to the Kerr model.

Another well-known problem related to the Kerr model is the instability that occurs near the inner Cauchy horizon, where an infinite amount of energy might accumulate, forming in fact a new space-time singularity, although the tidal forces can be finite, as opposed to the case of the central singularity \cite{Poisson_Israel_19890,Poisson_Israel_1989,Ori}. This problem is known under the name of "mass inflation".

The experimental data may help, in the near future, to better understand the nature of the sources of gravitational waves that we are able to detect.
The ringdown waveform of a black hole is 
completely 
determined by Quasinormal modes (QNMs), which depend only on the mass and angular momentum of the BH. Thus, every deviation from the standard result of GR may be associated to alternative theories of gravity or a different nature of the source with respect to the case of Kerr BH. In this respect, 
the possible presence of additional ``echoes'' 
in the ringdown waveform
has been largely debated in the last years 
(see the exhaustive review in Ref.\cite{ECO}) and alternatives to the BHs as   
gravastars~\cite{grav, grav2}, bosonstars~\cite{bs}, or other exotic compact objects ~\cite{Mark, ECO1, ECO2, ECO3, ECO4, ECO5}) have been investigated.

In this paper, we mainly study a class of non-singular topological BHs in $F(R)$-gravity, where the usual Einstein-Hilbert term in the gravitational Lagrangian is replaced by a smooth function of the Ricci scalar $R$. We are able to find the metric in analytic form and we can show that these black holes usually have two horizons, the inner one being of Cauchy type. However, the inner part of the black hole is a singularity-free region.

As mentioned above, a inner Cauchy horizon can trigger the mass inflation problem. It is not clear, a priori, whether this occurs also in regular black holes though. This issue has been recently investigated in regular  black holes with a de Sitter core \cite{Bonanno}. However, our solutions do not have this internal structure so we need to analyze the mass inflation problem again.

The structure of the paper is the following: in Sec. {\bf 2} we derive the analytic form of the non-singular black holes in $F(R)$-gravity and we study some of their properties. In Sec. {\bf 3} we investigate the problem of mass inflation for these solutions and we draw some conclusion in Sec. {\bf 4}.

%%%%%%%%%%%%%%%%%%%%%%%%%%%
\section{Non-singular black holes in $F(R)$ modified gravity} 

In this Section, following Refs. \cite{Dup,calza}, we present an exact and new class of non-singular vacuum topological black hole solutions within a particular  class of modified $F(R)$-gravity.

To begin with let us write down the equations of motion for $F(R)$-gravity \cite{faraoni,odi, Defe,capo, Seba, add0, add} in vacuum, namely with vanishing  stress tensor matter (here, $F_R(R)= d F(R)/dR$),
\begin{equation}
F_R(R) \left(
R_{\mu\nu}-\frac{1}{2}g_{\mu\nu}R
\right)
=\frac{1}{2}g_{\mu\nu}\left(F(R)- RF_R(R) \right)+\left(\nabla_\mu  \nabla_\nu-g_{\mu \nu} \nabla^2 \right)F_R(R)\,.
\label{fr}
\end{equation}

Let us consider the class of modified gravity models such that $F(0)=0$ and $F_R(0)=0$.
As a result, taking into account (\ref{fr}), any spherically symmetric static (SSS) metric (the ones we are interested in) such that leads to $R=0$ is a solution of the above equations of motion.

The most simple and important example is the scale invariant quadratic gravity $F(R)=R^2$, investigated in Ref.\cite{Max,Cognola:2015uva}, and its effective one-loop correction, investigated in Ref. \cite{max2,Rinaldi:2014gha}, 
\begin{equation}
F(R)=R^2-bR^2\ln\left( \frac{R^2}{\mu^2} \right) \,,
\label{fr1}
\end{equation}
where $b$ is a parameter of the model and $\mu$ a mass-scale.

Other examples are modifications of GR induced by one-loop corrections, for instance
\begin{equation}
F(R)=R-R_0\ln\left(1+\frac{R}{R_0} \right) \,.
\label{fr11}
\end{equation}
For very large $R_0$ this model reduces to quadratic gravity plus higher order term in Ricci scalar. For small $R_0$, this is GR plus small corrections.
For other examples, see Ref. \cite{calza}.

It is worth mentioning that 
models
without the linear term in $F(R)$ may suffer from instabilities
of 
Minkowski space-time against gravitational perturbations, since they might grow unbounded. Thus, in the following it is understood that our considerations are made in relations to $F(R)$ models with a linear term in $R$, like in eq. \eqref{fr11}.

Now let us look for a quite general SSS metric of the kind 
\begin{equation}
ds^2=-A(r)dt^2+\frac{ r dr^2}{(r-\ell)A(r)}+ r^2 dS_k^2\,,
\label{r1}
\end{equation}
in which $dS_k^2$ is the metric of two dimensional sphere $S^2$, two dimensional torus $T^2$ or two dimensional compact hyperbolic manifold $H^2/\Gamma$. Thus, we are investigating the existence topological black hole (for the case of GR see for example Ref. \cite{Vanzo,Birmingham:1998nr}).

The associated Ricci scalar is
\begin{equation}
R=\frac{1}{2 r^2}\left(2(\ell r-r^2)A''+ (7\ell-8r)A'-4(A-k) \right)\,.
\label{ricci}
\end{equation}
Imposing $R=0$, one gets the linear second order differential equation  
\begin{equation}
2(\ell r-r^2)A''+ (7\ell-8r)A'-4(A-k)=0\,.
\label{r3}
\end{equation}
The solution is
\begin{equation}
 A(r)=k-\frac{C}{r}-\frac{3C \ell}{r^{5/2}}\left( \sqrt{r-\ell}\ln(\sqrt{r-\ell}+\sqrt{r})-\sqrt{r} \right)+
   \frac{Q\sqrt{r-\ell}}{r^{5/2}} \,,
\label{r33}
\end{equation}
where $C$ and $Q$ are two constants of integration. In the following we make the choice $C>0$, keeping  $ Q$  real.  The horizons of the metric (\ref{r1}) are the non-negative real  zeros of $A(r)\left(\frac{r-\ell}{r}\right)$. There exists the trivial zero $r=\ell$ and the possible zeroes related to $A(r)$.

In order to discuss only the black hole case, we assume $\ell << C$, namely a microscopic length. In fact, if $\ell$ is sufficiently large, bigger than the zeroes of $A(r)$, one is dealing with a static wormhole (see for example \cite{ECO5}).  
In general $A(r)$ may have no zeroes, one or two real zeroes, as discussed below. However, one may note that when $\ell=0$, one has the same  Reissner-Nordstrom (RN)-like solution  found in \cite{Max}. For the case $k=1$ we have,
\begin{equation}
 A(r)=1-\frac{C}{r}+\frac{Q}{r^{2}} \,.
\label{r333}
\end{equation}
The associated horizon are the zeroes of $r^2-Cr+Q=0$, namely
\begin{equation}
r_{\pm}=\frac{C\pm \sqrt{C^2-4Q}}{2}\,.    \end{equation}
If $Q<0$, there is only a positive zero, $r_{+}=\frac{C+ \sqrt{C^2+4|Q|}}{2}$, which defines the event horizon. If $Q>0$, as in GR and RN black hole, and $ C^2> 4Q$, there are two positive zeroes. Thus, when $Q>0$, there exists also an inner or Cauchy horizon located at $r_{-}=\frac{C- \sqrt{C^2+4|Q|}}{2}$. 
In this case, as we already  have mentioned within  GR, the presence of Cauchy horizons may be  problematic and is related to the so called mass inflation.  
Finally, for $C^2< 4Q$, no horizon is present. We recall that these solutions are valid for any $F(R)$-gravity that satisfies $F(0)=F_R(0)=0$, thus extending the results found in \cite{Max} for $\ell=0$.

When $\ell$ is not vanishing, but sufficiently small, the situation is quite different, since $A(\ell)=1+\frac{2C}{\ell}>0$. Thus, for $C>0$, a Cauchy horizon seems to be always present and we numerically confirm this fact in the next subsection.
Moreover, the presence of a fundamental length scale in the solution requires $r\geq l$ and makes the metric singularity free. In \S \ref{Krus} chapter we will examine the possibility to extend the solution beyond the boundary at $r=l$.

\subsection{Determination of the horizons and extendibility of the solution \label{hh}}

\begin{figure}[H]
  \centering 
   \includegraphics[scale=0.5]{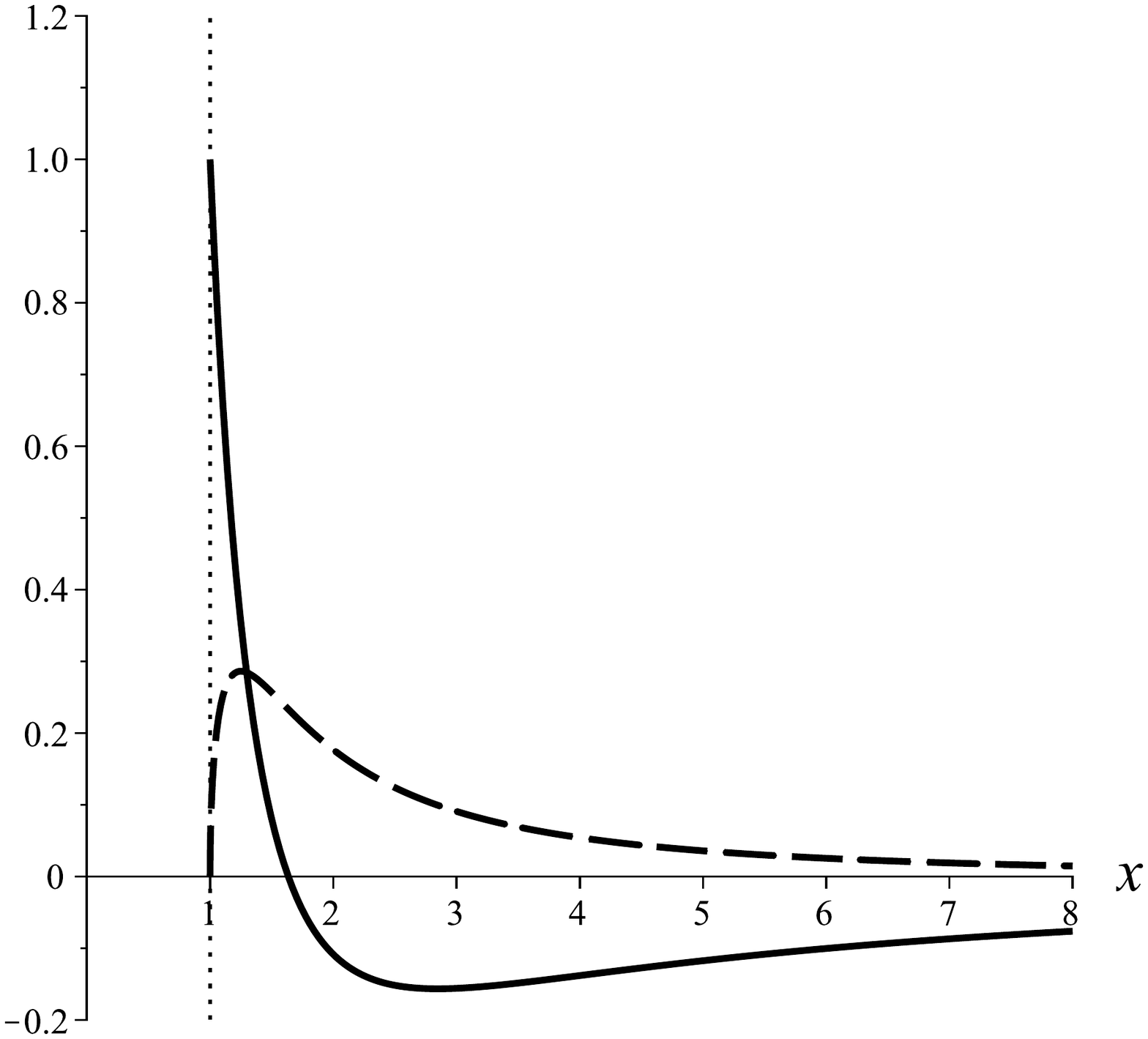} 
  \caption{Plot of $B_{1}$ (solid line) and $B_{2}$ (dashed line). The vertical line denotes the boundary of the spacetime at $x=1$. Note that both functions are finite at $x=1$.}
    \label{fig1}
  \end{figure}

In this chapter, for the sake of simplicity, we will assume $k=1$, namely the spherical topology in (\ref{r1})--(\ref{r33}).
In order to  study numerically the presence of horizons, we set $x=r/\ell$ and $B(x)=A(x)-1$ so that Eq. (\ref{r3}) can be written as the standard hypergeometric equation
\bea\label{Beq}
x(1-x)B''+\left[\gamma-(\alpha+\beta+1)x\right]B'-\alpha\beta B=0\,,
\eea
with $(\alpha,\beta,\gamma)=(1,2,7/2)$ or  $(\alpha,\beta,\gamma)=(2,1,7/2)$. For the first set of coefficients, the solutions of the equation around the point $x=1$ are given in terms of the hypergeometric functions \cite{nist},
\bea
B_{1}&=&F(\alpha,\beta,1+\alpha+\beta-\gamma;1-x)=F\left(1,2,\frac12;1-x\right)\,,\\\non\\\non
B_{2}&=&(1-x)^{\gamma-\alpha-\beta}F(\gamma-\alpha,\gamma-\beta,1+\gamma-\alpha-\beta;1-x)=F\left(\frac32,\frac52,\frac32;1-x\right)\,.
\eea
The second set of coefficients gives the same solutions, due to the symmetry properties of the hypergeometric functions.
In terms of elementary functions, they read
\bea
B_{1}&=&-\frac{1}{2}\frac{(x-3)}{ x^{2}}-\frac{3}{2} \frac{\sqrt{x-1}\arcsinh(\sqrt{x-1})}{ x^{5/2}}\,,\\\non\\\non
B_{2}&=&\frac{\sqrt{x-1}}{ x^{5/2}}\,,
\eea
thus the general solution to eq.\ \eqref{Beq} is any linear combination of these functions.
Asymptotically, these functions have the following properties
\bea
\lim_{x\rightarrow\infty}B_{1,2}(x)=0\,,\quad \lim_{x\rightarrow1^{+}}B_{1}(x)=1\,,\quad  \lim_{x\rightarrow1^{+}}B_{2}(x)=0\,.
\eea
The function $B_{1}$ has only one zero at $x\simeq 1.63491$ and a global minimum at a point in the interval $[1,2]$. The unique zero of $B_{2}$ is instead at $x=1$ and there is a global maximum at a point in the interval $[1,2]$, see fig.\ \eqref{fig1}.

As physical solution, we seek the one that matches the usual Schwarzschild metric for large distances. Thus, we choose the combination 
\bea\label{effA}
A=1-\frac{2m\gamma(x)}{ r}\,,
\eea 
where $m>0$ is an arbitrary constant with mass dimension and $\gamma(x)=-2xB_1(x)$. In the large $x=r/ \ell$ limit  we have 
\bea
\gamma(x)\simeq 1+{3\over x}\left[\ln( 2\sqrt{x})-1\right]+{\cal O}\left({1\over x^2}\right)\,,
\eea 
thus the line element rapidly converges to the Schwarzschild solution if $\ell$ is taken to be of planckian size and $m$ of the order of the mass of the Sun.
If we add the $B_2$ solution, we have a new term  that asymptotically behaves like $\ell^2/r^2$, giving rise to a trivial deformation of the exact Reissner-Nordstrom-like  black hole solution found in the $R^2$-theory \cite{Max}.

One can prove that the curvature invariants are finite for all $r\geq \ell$. In particular
\bea
\lim_{r\rightarrow\ell^+}R_{\mu\nu}R^{\mu\nu}={3\ell^2-8m\ell+48m^2\over 2\ell^2}\,,\quad \lim_{r\rightarrow\ell^+} R_{\alpha\beta\mu\nu}R^{\alpha\beta\mu\nu}={6\ell^2+12m\ell+96m^2\over 2\ell^6}\,,
\eea 
thus the spacetime is free of curvature singularities.

\begin{figure}[ht]
  \centering 
   \includegraphics[scale=0.5]{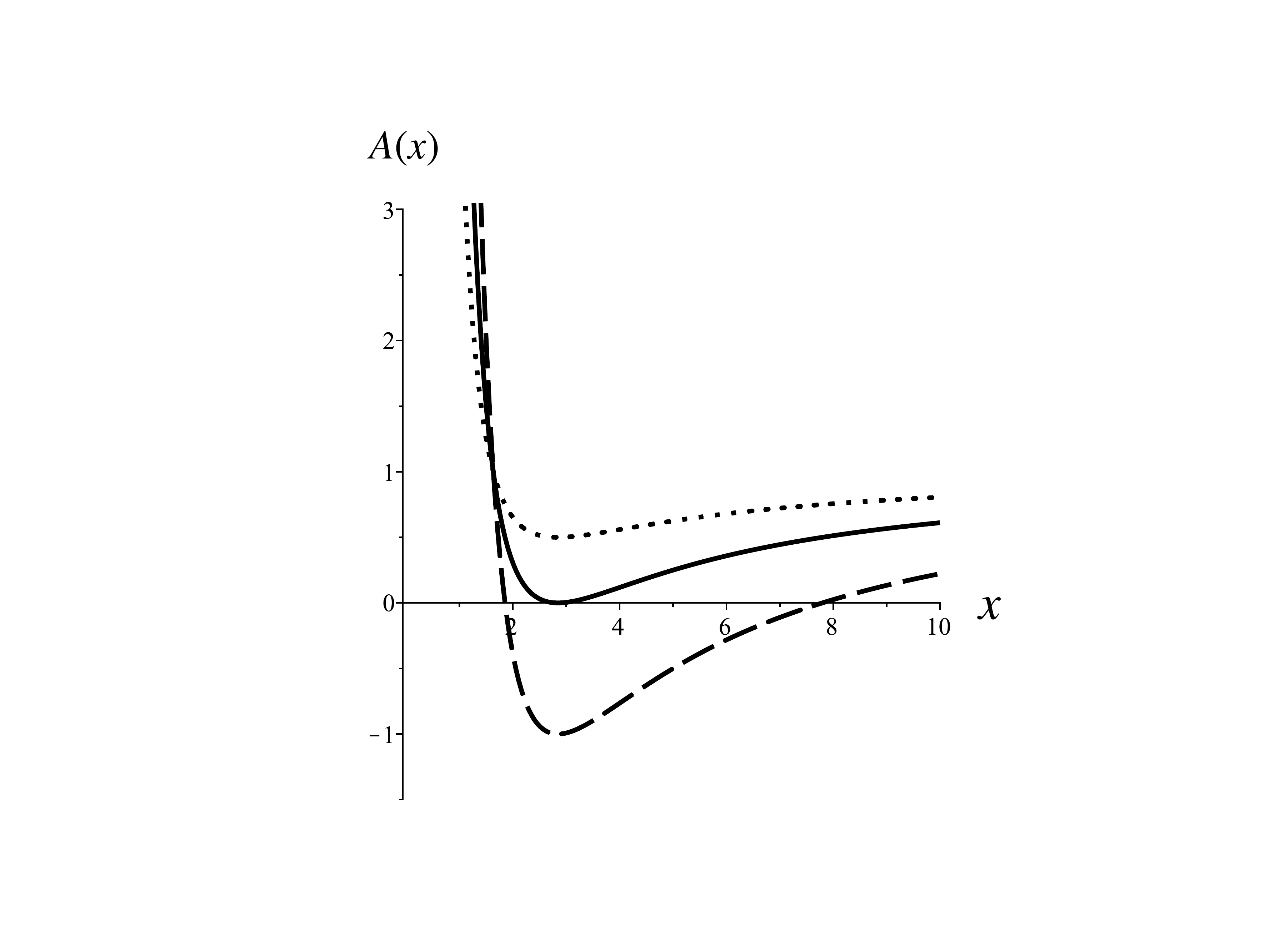} 
  \caption{Plot of $A(x)$ for $\rho=\rho_{\rm cr}$ (solid line), $\rho=2\rho_{\rm cr}$ (dashed line), $\rho=\frac12 \rho_{\rm cr}$ (dotted line).}
    \label{fig2}
  \end{figure}

One can numerically check that the function $A(x)$ has a global minimum at $x_{\rm min}\simeq 2.846$ provided $m>0$. It follows that there exists a critical ratio $\rho_{cr}=m/\ell\simeq 1.596$ such that, for $\rho>\rho_{\rm cr}$ there are two event horizons, which coincides when $\rho=\rho_{\rm cr}$, while for $\rho<\rho_{\rm cr}$ there are none, see fig. (\ref{fig2}). Note that the horizons, when they exists, are always outside the limiting surface $x=1$. 

By assuming that $\ell$ is a microscopic fundamental length, astrophysical black holes are in the regime $\rho\gg \rho_{\rm cr}$ so they all have two horizons, the outer one being almost coincident with the standard Schwarzschild horizon (since $-2xB_1\simeq 1$).

\subsection{Motion of test particles\label{Krus}}

In this subsection we study the motion of both massless and massive test particles on the equatorial plane (i.e. with $\theta=\pi/2$) of the geometry described above. We follow the standard approach by using the Killing vectors of the metric
\bea
K^{\mu}=(1,0,0,0)\,,\qquad R^{\mu}=(0,0,0,1)\,,
\eea 
and the associated conserved quantities (energy and angular momentum)
\bea
E=-K_\mu{dx^\mu\over d\lambda}\,,\qquad L=r^2{d\phi\over d\lambda}\,,
\eea 
where $\lambda$ is an affine parameter along the geodesic path of the particle. Let us further define the quantity
\bea
\varepsilon=-g_{\mu\nu}{dx^\mu\over d\lambda}{dx^\nu\over d\lambda}\,,
\eea 
such that $\varepsilon=0$ for massless particles and $\varepsilon=1$ for massive ones.

By combining these expressions we find 
\bea
{E^2\over 2}={A(r)\over 2}\left(\varepsilon+{L^2\over r^2}\right)+\frac12 {r\over r-\ell}\left(dr\over d\lambda\right)^2\,.
\eea 
Upon the rescaling\footnote{The constant of integration is chosen so that $r=\ell$ implies $\tilde r=\ell$.} 
\bea
\tilde r=\int dr\sqrt{r\over r-\ell}=\ell-{\ell\over 2}\ln \ell+\sqrt{r(r-\ell)}+{\ell\over 2}\ln\left[2r-\ell+2\sqrt{r(r-\ell)}\right]\,,
\eea
we find the standard form
\bea\label{modpot}
{E^2\over 2}={A(\tilde r)\over 2}\left(\varepsilon+{L^2\over r(\tilde r)^2}\right)+\frac12 \left(d\tilde r\over d\lambda\right)^2\,,
\eea 
which mimics the motion of a particle in the potential
\bea\label{pot_orbit}
V={A(\tilde r)\over 2}\left(\varepsilon+{L^2\over r(\tilde r)^2}\right)\,,
\eea 
with a constant total energy.

\begin{figure}[H]
  \centering 
   \includegraphics[scale=0.55]{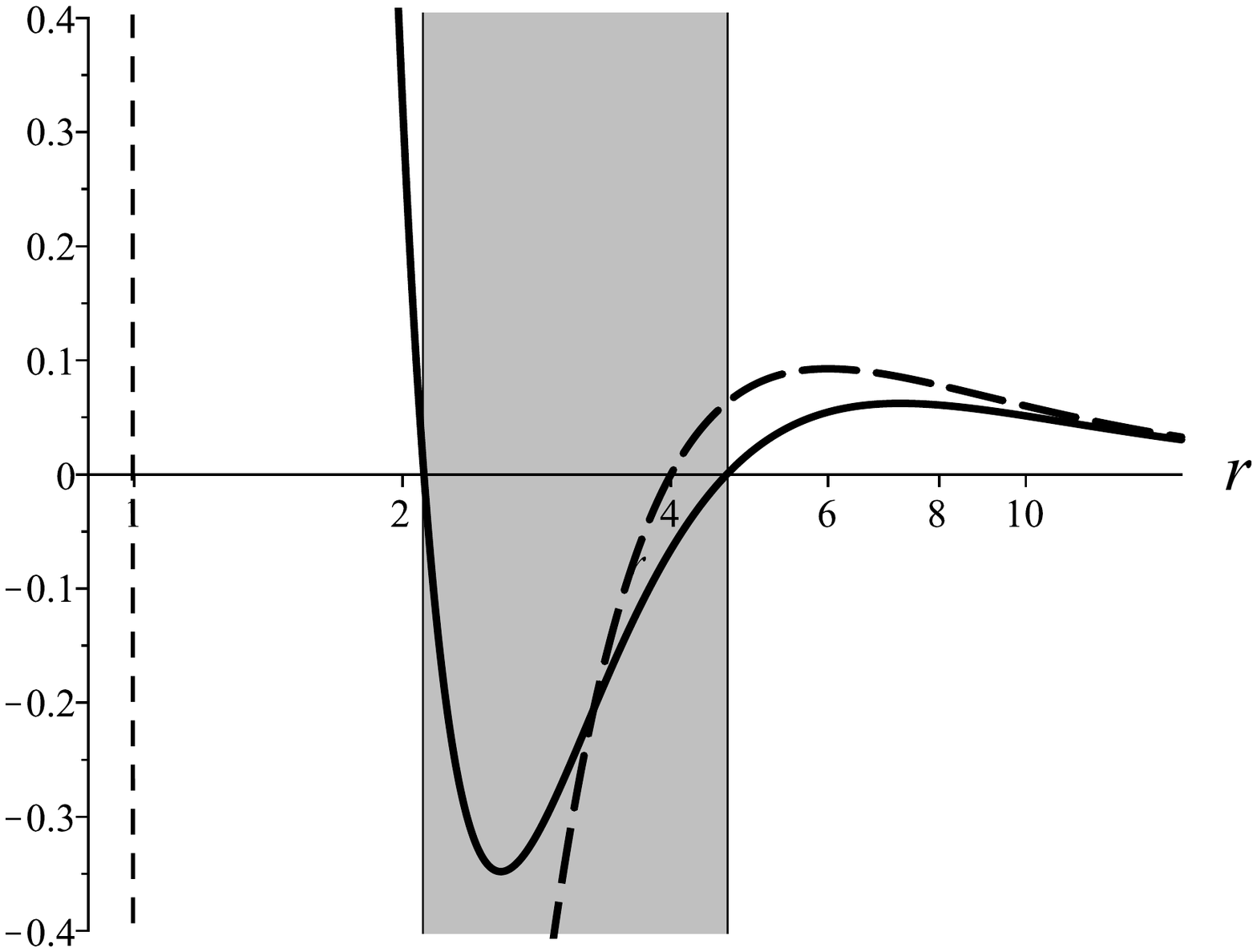} 
  \caption{Comparison between the Schwarzschild potential (dashed curve) and the potential $V/L^2$ (solid line) of eq. \eqref{pot_orbit}  for $\epsilon=0$ and $m=1.7\ell$. The shaded area marks the region between the two horizons. The dashed vertical line marks the boundary of the spacetime at $r=\tilde r=1$.}
    \label{massless_plot}
  \end{figure} 

Let us first consider the motion of a massless particle (i.e. $\varepsilon=0$). In the Schwarzschild case, the potential has an absolute maximum located at $r=3m$ for all $L$, which coincides with the photon sphere of the black hole. In our case, there are both a local maximum and a local minimum. It can be proved numerically that the former is displaced at larger distances than $r=3m$ by a quantity that rapidly converges to zero as $\ell\rightarrow 0$. The latter is located in between the two horizons (we are considering the $\rho>\rho_{\rm cr}$ case). The qualitative behaviour of the potential is plotted in fig. \ref{massless_plot} where the standard Schwarzschild case is also showed for comparison\footnote{The plot represents the curve $V(r)/L^2$ instead of $V(\tilde r)/L^2$. However, the function $r(\tilde r)$ is monotonic and growing in the range of interest, so the plot is qualitatively the same.}. From this qualitative picture we can draw some considerations. In the Schwarzschild geometry, a photon can be scattered by the potential, it can plunge into the horizon, or it can orbit for some time on the photon sphere. In our case the phenomenology is much richer. Photons with low energy compared to the local maximum of the potential are scattered back to infinity. However, there exists a family of stable orbits within the two horizons, where photons can eventually tunnel from the outside. Among these, the circular orbit corresponds to the global minimum of the potential. High energy photons that classically cross the outer horizon  cross also the inner horizon. They can be scattered by the potential back into the region between the horizons escape to infinity if their energy is lower that the value of the potential at $\tilde r=1$. If their energy is higher than this value photons can reach the boundary of spacetime at $\tilde r=1$.

\begin{figure}[H]
  \centering 
   \includegraphics[scale=0.55]{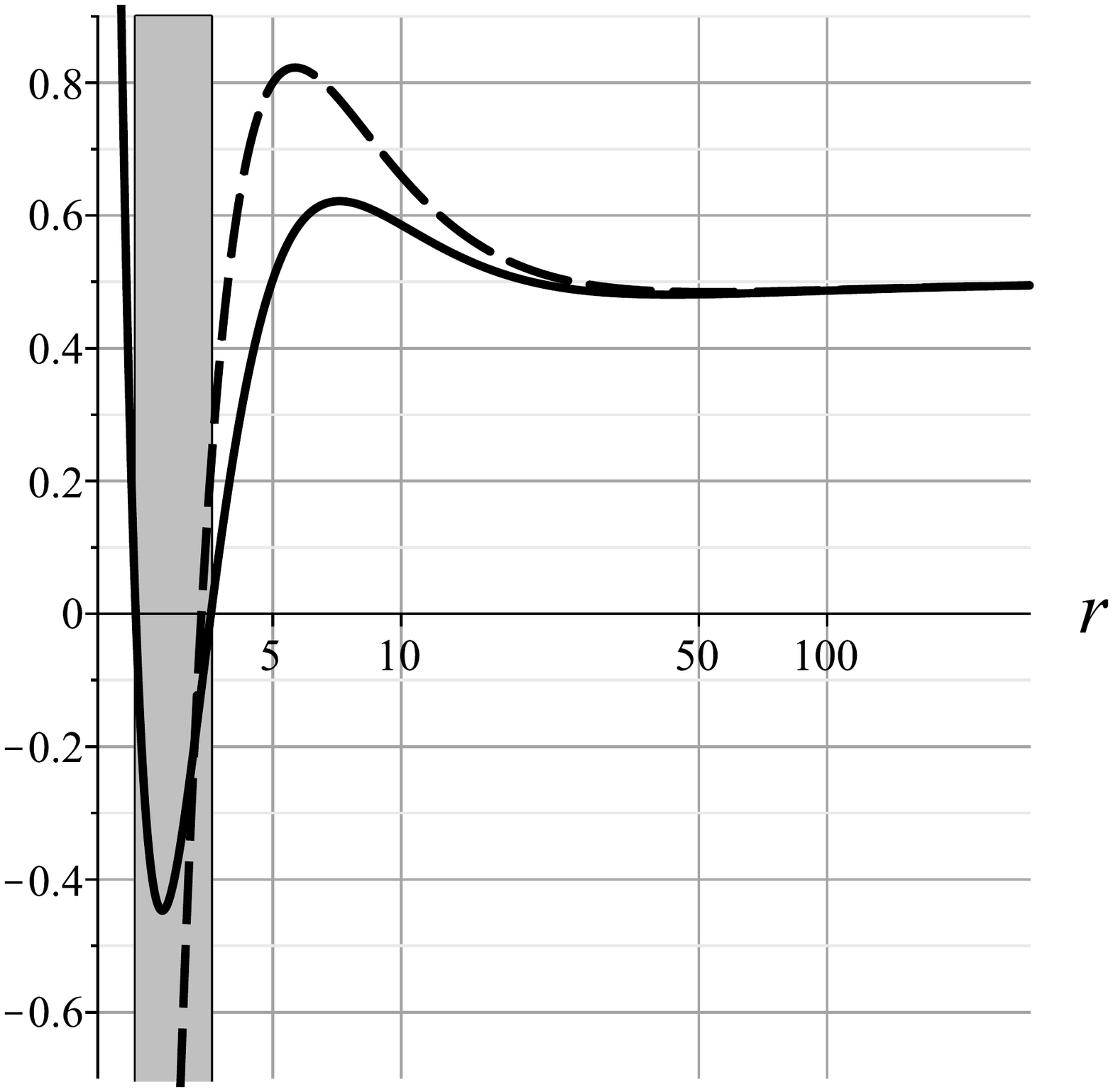} 
  \caption{Comparison between the Schwarzschild potential (dashed line) and the potential $V$ of eq. \eqref{modpot} (solid line) for $\epsilon=1$, $L^2=10$, and $m=1.7\ell$. The shaded area denotes the space between the horizons.}
    \label{mass_plot}
  \end{figure}

The case of a massive test particle (i.e. $\epsilon=1$) is pretty much alike. As shown in fig. \ref{mass_plot}, the potential is very similar to the standard one outside the outer horizon. Thus, according to the value of the angular momentum $L$ there can be a maximum and a minimum (hence stable orbits) or no extremal point (so the particle directly plunges into the horizon). As before, the position and the height of the extreme points is modified by a quantity that vanishes rapidly when $\ell\rightarrow 0$. Between the two horizons, in contrast with the Schwarzschild case, the potential has a global minimum, opening the possibility of stable orbits. Moreover, as in the massless case, sufficiently energetic particles can cross the inner horizon and reach the boundary of the spacetime at $\tilde r=r=1$ or be bounced back  by the potential towards spatial infinity.

\begin{figure}[ht]
  \centering 
   \includegraphics[scale=0.5]{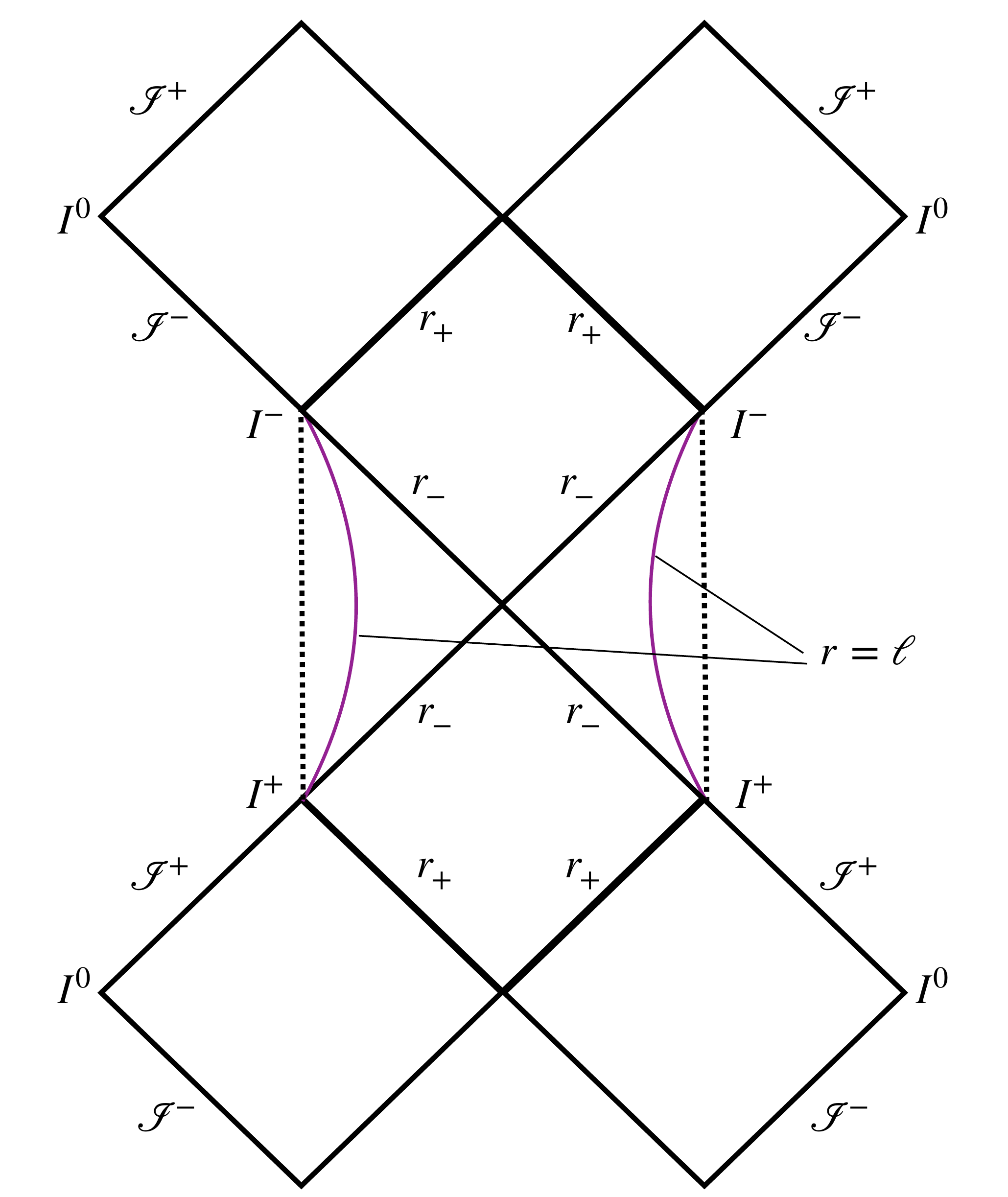} 
  \caption{Penrose diagram of the metric \eqref{r1}. It is very similar to the one of a RN black hole, except that the singularity (dashed vertical line) is now covered by the region delimited by $r=\ell$ (purple lines).}
    \label{penrose}
  \end{figure}

The global spacetime geometry of the solution can be determined in the following way. The function $A$ has two distinct zeros at $r_\pm$, corresponding to the two horizons found above. In addition, when $r=\ell$, $A= 1+4m/\ell$. Then, one can define the usual tortoise coordinate $r^*$ by integrating
\begin{equation}
dr^*=\sqrt{\frac{r}{r-l}}\frac{d r}{A(r)}\,,
\label{rstar}
\end{equation}
and the null coordinates 
\begin{equation}
v=t+r^*\,,\quad u=t-r^*\,.  \label{uv} 
\end{equation}
In terms of the latter, the metric reads
\begin{equation}
ds^2=-A(r)dvdu+r^2dS_1^2\,. 
\end{equation}
To study the maximal extension of this metric, one can proceed exactly like in the case the Reissner-Nordstr\"{o}m (RN) black hole because the metric has two horizons, located at $r_\pm$ (the actual value of $r_\pm$ is irrelevant for what it follows). For $r>r_+$ one defines the new null coordinates
\begin{equation}
    U^+=-e^{-k_+u}\,,\quad V^+=e^{k_+v}\,,
    \label{UVplus}
    \end{equation}
where $k_+$ is a constant, possibly related to the surface gravity at the external horizon. The metric becomes
\begin{equation}
   ds^2=-{A(r)\over k_+^2}\,e^{-2k_+r^*}dU^+dV^++r^2dS_1^2\,. 
\end{equation}
By expanding $A$ around $r_+$ and by integrating \eqref{rstar}, one can show that either $U^+=0$ or $V^+=0$ corresponds to the surface $r=r_+$, where $r^*\rightarrow -\infty$. Then, even though $U^+<0$ and $V^+>0$ according to the definition  \eqref{UVplus}, we see that we can extend these coordinates also in the region $r_-<r<r_+$, where $U^+>0$ or $V^+<0$. Then, by following the same prescriptions for the RN black hole, we can extend further the region defined by $U^+>0$ and $V^+>0$ by adopting the new null coordinates
\begin{equation}
    U^-=-e^{k_-u}\,,\quad V^-=-e^{k_-v}\,,
\end{equation}
where, again, $k_-$ is a constant related to the surface gravity at $r_-$. Then, the region defined by $U^-<0$ and $V^-<0$ connects smoothly to the one defined by $U^+>0$ and $V^+>0$. As before, $U^-$ and $V^-$ can be analytically extended to $U^->0$ and $V^->0$ through the surface $r=r_-$. This region, in terms of $r$, extends all the way down to $r=\ell$. This is where our solution differs form the RN one, in that there is no timelike singularity since the latter is cut away by the wall at $r=\ell$. To clarify further the structure of spacetime, and by using standard coordinate changes, one can depict the metric using the Penrose diagram as in fig. \ref{penrose}.

One might wonder what happens when a particle hits the wall at $r=\ell$. We have seen above that the potential energy is proportional to $A$ (for simplicity we consider a head-on collision with a massive particle so $\epsilon=1$ and $L=0$ in \eqref{pot_orbit}). Thus, when $r\sim \ell$, $V\sim 1+4m/ \ell$. If we assume that $\ell$ is of the order of the Planck size and $m$ of the order of few solar masses, we see that the (repulsive) potential wall is huge, of the order of $m/\ell\sim 10^{39}$ in natural units. Therefore, to effectively hit the wall, the incoming particle should have a kinetic energy so large that it would no longer be a test particle. Instead, it would become a source for the modified Einstein equations, making the solution \eqref{effA} no longer valid.

% Tesi Bertipagani

\section{Mass inflation}

As it is well known, the presence of the Cauchy horizon may lead to some strong instability. At the origin of such instability there is the exponential growth of 
the mass parameter of the solution  under perturbations caused by a crossflow of infalling and outgoing radiation near to the Cauchy horizon. The problem of the so-called mass inflation was firstly investigated by Poisson and Israel in Refs. \cite{Poisson_Israel_19890, Poisson_Israel_1989} in relation to the Kerr and Reissner-Nordstrom solutions, modelled with Vaidya spacetimes, in the framework of GR (see also the seminal work of Ori in Ref. \cite{Ori}). The Vaidya space-time describes the non-empty spherical symmetric charged black hole solution with a pure flux of radiation, where an inner Cauchy horizon is present.  

Recently, Bonanno et al. pointed out that if the singularity at $r=0$ is removed, as it happens with regular black holes with a de Sitter core, the mass inflation problem may be avoided \cite{Bonanno}. In this section we will investigate the issue of mass inflation in our class of non-singular black hole solutions in $F(R)$-gravity, which differ from the ones considered in \cite{Bonanno} by lacking a de Sitter core.

Some important remarks are in order. As we have seen, in $F(R)$-gravity the non-singular solutions can be obtained in vacuum, without invoking the presence of exotic matter, as opposed to the case of regular black hole solutions in GR, where matter with negative equation of state parameter is required \cite{BHR1, BHR2, BHR3, BHR4, BHR5, BHR6, BHR7}. 
On the other hand, the perturbation is generated by a radiation flux but in our treatment we will investigate the behaviour of the metric only, without a direct analysis of the stress-energy tensor. The procedure holds true as long we can identify the BH mass with
one thermodynamic parameter of the solution. 
In principle, in $F(R)$-gravity four integration constants may be present due to the fact that the field equations of the theory are fourth-order differential equations. 
Thus, given the metric (\ref{r1}), we will assume that $l$ is fixed and is a fundamental length scale of the solution, while 
\begin{equation}
A(r)\equiv A(r; m)\,,
\label{presc}
\end{equation}
where $m$ is the mass of the black hole and in principle can be a combination of constants. For example, in the case of the solution (\ref{r33}), the mass is not necessarily identified with $C$, since also $Q$ is present. This is an important difference with respect to the GR case and in principle we must not expect to find the same results using the same metrics.
For a discussion about black hole thermodynamic in $F(R)$-gravity see Refs. \cite{EI1, EI2, EI3, EI4, Seba1, Seba2, Seba3}. 

In contrast to the original work \cite{Poisson_Israel_1989}, we make an important simplification, modeling the infalling and outgoing radiation as thin shells, instead of a continuum. This allows to apply a generalized form of the Dray-'t Hooft-Redmount (DTR) relation \cite{Barrabes_Israel_1991}. 

To describe the null geodesic, it is convenient to introduce the advanced and retarded time coordinates $u\,,v$ as in (\ref{uv})--(\ref{rstar}).
Thus, the metric (\ref{r1}), in terms of the new coordinates, reads,
\begin{equation}
    ds^2=-A(r; m)dw^2\pm2\sqrt{\frac{r}{r-l}}dw dr +r^2dS^2_k\,,\label{r2}
\end{equation}
with $w=v\,,u$, respectively.

\begin{figure}[h]
    \centering
    \includegraphics[scale=0.65]{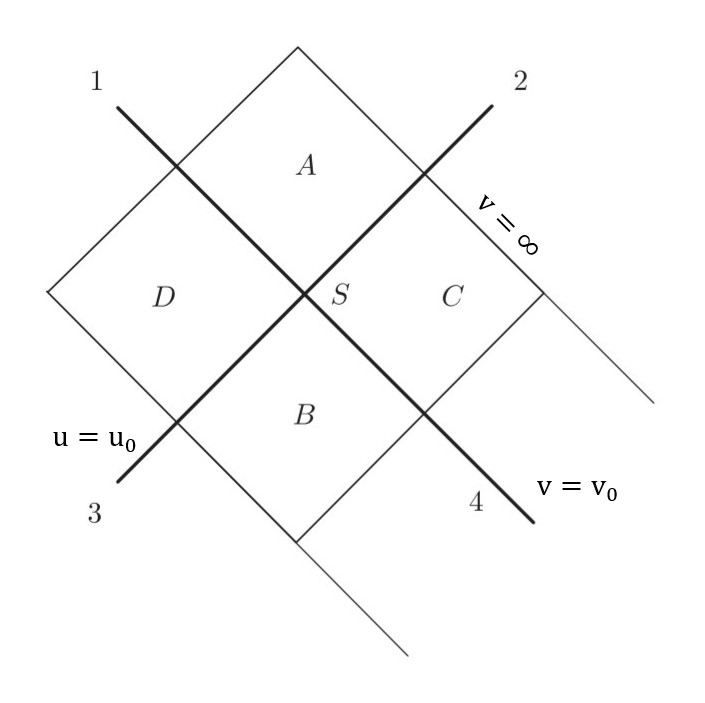}
    \caption{An infalling and an outgoing thin null shells collides at the two sphere $S$.}
    \label{fig8}
\end{figure}

We consider an infalling and an outgoing thin shell of radiation, described respectively by the equations $v=v_0$, $u=u_0$, which collide in the region between the event and the Cauchy horizon at the two-sphere $S=(u_0,v_0)$, 
splitting the space-time into four sectors, as shown in fig. \ref{fig8}. We assume that no singularities arise where the shells collide \cite{Barrabes_Israel_1991} and that
each sector is described by the same metric but with a different mass parameter $m_A\,, m_B\,, m_C\,, m_D$.
In particular, $m_C-m_B$ is the energy of the infalling radiation and $m_B-m_D$ is the energy of the outgoing radiation. 

The mass parameter $m_B$ corresponds to the mass of the black hole measured by an external observer before the infalling shell crosses the event horizon, such that is natural to choose $v$ to be the advanced time defined in sector $B$. As a consequence, the Cauchy horizon is located at $v=\infty$ and corresponds to the value $r=r_{-}$ (namely, $r^*\rightarrow+\infty$) for which $A(r_{-}; m_B)=0$.

The generalized DTR relation
is a geometric condition that allows us to relate the metrics in each of
these regions to each other, evaluated on the collision two-sphere $S$. In general, they are valid provided the proper junction conditions through the hypersurface separating the shells are met. 
These conditions
can be derived from integrating the field equations across the hypersurfaces. In
 GR, they can be rewritten in terms of the stress-energy tensor.
In $F(R)$-gravity this operation is not so trivial apart for the case of pure quadratic gravity with $F(R)=R^2$ \cite{Senovilla}. However, in Ref. \cite{Mann}, Brown et al.  suggested a way to derive the DTR relations without the direct use of the field equations and in what follows we will briefly recall their argument.
  
A hypersurface $\Sigma$ is null if its normal vector 
$l_\mu$ concides with the vector tanget to its generators and satisfies the relation
\begin{equation}
g^{\mu\nu}l_\mu l_\nu=0\,,    
\end{equation}
together with the geodesic equation
\begin{equation}
l^\nu\nabla_\nu l^\mu=\kappa l^\mu\,,
\end{equation}
where $\kappa$ is the `acceleration' on the shell.
Therefore, the generators of a null hypersurface are null geodesics parametrized by a non-affine parameter $\lambda$ such that $l^\mu=d x^\mu/d\lambda$. We can introduce the following coordinates
\begin{equation}
    y^i=(\lambda, \theta^i)\,,
\end{equation}
where $\theta^i$ are spatial coordinates labelling the generating geodesics. In particular the tangent vectors to $\Sigma$ are given by
\begin{equation}
e^\mu_i=\frac{d x^\mu}{d \theta_i}\,, \quad l_\mu e^\mu_i=0\,.    
\end{equation}
In the case of SSS space-time with spherical topology ($k=1$), we can identify $\theta^i$ with the angular coordinates, namely
$\theta_i=\theta,\phi$ and
$  e_\theta^\mu=\partial_\theta^\mu\,, e_\phi^\mu=\partial_\phi^\mu$.

The induced metric on the hypersurface is defined as
\begin{equation}
\sigma_{i j}=g_{\mu\nu}\frac{\partial x^\mu}{\partial y^i} \frac{\partial x^\nu}{\partial y^j}  =
g_{\mu\nu}\frac{\partial x^\mu}{\partial \theta^i} \frac{\partial x^\nu}{\partial \theta^j}\,,
\end{equation}
where one has to make use of the fact that $l_\mu l^\mu=0=l^\mu e_i^\mu$.
Finally, the extrinsec metric on $\Sigma$ reads,
\begin{equation}
K_{ij}=\frac{1}{2}\mathcal {L}_{l^\mu}   \sigma_{i j}\,,
\end{equation}
where $\mathcal {L}_{l^\mu}$ is the Lie derivative along the vector $l^\mu$.
%If we introduce an auxiliary null vector field $N^\mu$ such that $N_\mu l^\mu=-1\,,\, N_\mu e^\mu_i=0$, we can also write the completeness relation for the inverse metric $g^{\mu\nu}$,
%\begin{equation}
%g^{\mu\nu}=-2^{(\mu}N^{\nu)}+\sigma^{ij}e_i^\mu e_j^\nu\,.    
%\end{equation}
Let us come back to fig. \ref{fig8}.
We have
four null hypersurfaces $\Sigma_{1,2,3,4}$ which correspond to the null thin shell with their relative sets of coordinates and vectors tangent to their generators $l_{1,2,3,4}$ depending on the non-affine parameters $\lambda_{1,2,3,4}$. 
The trace of the extrinsec metric, 
$K=\sigma_{ij}K^{ij}$, corresponds to the extrinsec scalar curvature and the extrinsec scalar curvatures $K_{1,2,3,4}$ of the  null surfaces describe the expansions of a congruence of null geodesics, not affinely parametrized, orthogonal to the surfaces. Due to the assumptions made at the beginning of this Section, only two independent null directions perpendicular to each surface are present and we can write the following geometric condition, which holds on the two-sphere $S$:
\begin{equation}
 (l_1 \cdot l_2 )(l_3\cdot l_4) =(l_1\cdot l_4)(l_2\cdot l_3)\,.
\end{equation}
This equation corresponds to what is called the DTR relation, namely
\begin{equation}\label{eq109}
    |Z_AZ_B|=|Z_CZ_D|\,,
\end{equation}
where the scalar functions $Z_{A,B,C,D}$ depend on the expnaions of the shells, namely the extrinsec curvatures. For $Z_A$ we have
\begin{equation}
Z_A=\frac{K_1 K_2}{l_1 l_2}\,, 
\end{equation}
and similar expressions are given for the other sectors. We note that this scalar functions are independent on the choice of the non-affine parameters used to parametrize the generators. 

The DTR relation
connects the metric in the four space-time sector on the two-sphere $S$ in a way that is independent of the parameters
used to generate the shells. Moreover, in this derivation the field equations of the theory have been not used and we may see the DTR relation as 
a necessary geometric condition of the solution itself.

In Ref. \cite{Poisson_Israel_1989} it is shown that for a metric in the form of (\ref{r1}) the DTR relation reads,
\begin{equation}\label{eq117}
    |A(r;m_A)A(r;m_B)|=|A(r;m_C)A(r;m_D)|\,,
\end{equation}
which expresses the relationship between the four masses $m_{A}\,,m_B\,,m_C\,,m_D$ on the two-sphere $S$.

In order to proceed further, two important assumptions are necessary:
\begin{itemize}
    \item $m_B\neq m_D$: this is equivalent to assume that there exists outgoing radiation;
    \item $m_C - m_B \sim v^{-p}\,,p\geq 12$: we assume that the Price law \cite{Prince} is valid in the model under consideration. This should be true if the backscattering that generates such behavior occurs far from the event horizon, where the metric must turn out to be, with good approximation, Schwarzschild. This is true in our case.
\end{itemize}

We begin by assuming $l\ll r_-$. This choice corresponds to the general case (\ref{r33})
with $l\ll C$ (for $k=1$).

%%%%%%%%%%%%%%%%%%%%

\subsection{The case $l\ll r_-$}

We begin by writing $A(r;m_B)$ near the Cauchy horizon, where $A(r_{-}; m_B)$ vanishes, in terms of the advanced time $v$. By making use of the Killing surface gravity evaluated on the horizon,
\begin{equation}
 \kappa_-=-\frac{1}{2}\sqrt{\frac{r_{-}-l}{r_{-}}}\frac{d A(r; m_B)}{d r}|_{r=r_{-}}\simeq 
 -\frac{1}{2}\frac{d A(r; m_B)}{d r}|_{r=r_{-}}>0\,,\label{kappa1}
\end{equation}
where we have taken into account that $l\ll r_-$,
one has
\begin{equation}\label{eq119}
    A(r;m_B)\simeq \frac{d A}{d r}(r;m_B)|_{r_-}(r-r_-)+O(r-r_-)\simeq -2\kappa_- (r-r_-)\,.
\end{equation}
By working in Eddington-Finkelstein coordinates (\ref{r2}) with $w=v$ and $d S_k^2=0$,
we get for radiation with $d s^2=0$, 
\begin{equation}\label{eq119a}
    \frac{dr}{dv}=\frac{1}{2} \sqrt{\frac{r-l}{r}}A(r; m_B)\,.
\end{equation}
In the limit $r\rightarrow r_-$ and $l\ll r_{-}$, by taking into account  (\ref{eq119}), we easily obtain
\begin{equation}\label{eq120}
    \frac{dr}{dv}\simeq -\kappa_-(r-r_-)\,,
\end{equation}
such that
\begin{equation}
A(r;m_B)\simeq -2\kappa_-e^{-\kappa_- v}\,,
\label{AA1}
\end{equation}
which correctly vanishes at the Cauchy horizon where $v\rightarrow +\infty$.
By using this result in Eq. \ref{eq117} we get,
\begin{equation}\label{eq121}
    |A(r;m_A)|=|A(r;m_C)A(r;m_D)|\frac{1}{2\kappa_-}e^{\kappa_- v}\,.
\end{equation}
The assumption $m_D\neq m_B$ implies that $A(r;m_D)\neq0$ in the limit $r\rightarrow r_-$.
On the other hand, $A(r; m_C)$ deserves more attention. 
By using the second assumption mentioned above, which leads to $m_C\sim m_B+v^{-p}$ ($p\geq 12$), in the limit $v\rightarrow +\infty$ and $l\ll r_{-}$ we find at the first order,
\begin{equation}\label{eq122}
    \begin{split}
    A(r;m_C)&\simeq A(r;m_B)+\frac{\partial A}{\partial (v^{-p})}\Bigr\rvert_{v^{-p}=0} v^{-p}\simeq
     \frac{\partial A}{\partial (v^{-p})}\Bigr\rvert_{v^{-p}=0} \ v^{-p} -2\kappa_-e^{-\kappa_- v}
    \end{split}\,.
\end{equation}
Inserting this expression into (\ref{eq121}), we have
\begin{equation}\label{eq123}
    |A(r;m_A)|\simeq\left\rvert-1+\frac{1}{2\kappa_-}\frac{\partial A}{\partial (v^{-p})}\Bigr\rvert_0 \ v^{-p}e^{\kappa_-v}\right\rvert|A(r;m_D)|\,.
\end{equation}
Thus, in the limit $v\rightarrow\infty$, we derive the following behaviour for $A(r; m_A)$,
\begin{equation}\label{eq124}
    |A(r;m_A)|\simeq v^{-p}e^{\kappa_- v}\,,
\end{equation}
and the metric becomes unbounded as the Cauchy horizon is approached.

\subsection{The case $l\sim r_-$ and the limit $l=r_-$}

Now we will check what happens in the limit $l\sim r_-$, i.e. when the Cauchy horizon has a size comparable with the minimal length of the solution. 
This choice corresponds to the solution (\ref{effA}), where we remind that in order to have two distinct horizons we have to require $1.596 \lesssim m/l$. A numerical check shows that the distance between the two horizons increases with the value of $m/l$. However, the Cauchy horizon results to be located near to the value $r_-\simeq 2l$. In particular, for $m/l\rightarrow \infty$, we observe that the inner horizon position tends to the value $r_-= 1.635$.

The generalized DTR relation (\ref{eq117}) and the assumptions on the mass parameters are independent of the value of $l$ 
but now the approximation in (\ref{kappa1}) is no more valid and the surface gravity on the Cauchy horizon reads
\begin{equation}\label{eq125}
     \kappa_-=-\frac{1}{2}\sqrt{\frac{r_{-}-l}{r_{-}}}\frac{d A(r; m_B)}{d r}|_{r=r_{-}}>0\,.
\end{equation}
Therefore, Eq. (\ref{eq119}) takes the form
\begin{equation}\label{eq126}
    A(r;m_B)\simeq  -2 \sqrt{\frac{r_-}{r_- -l}}\kappa_- (r-r_-)\,.
\end{equation}
Equations (\ref{eq119a})--(\ref{eq120}) are still valid in the limit $r\rightarrow r_{-}$ and 
in first approximation we obtain
\begin{equation}\label{eq127}
     A(r;m_B)\simeq -2\kappa_-\sqrt{\frac{r_-}{r_- -l}}e^{-\kappa_- v}\,.
\end{equation}
Following the same derivation as the previous section, we infer an approximate expression for $A(r;m_C)$,
\begin{equation}\label{eq128}
    \begin{split}
    A(r;m_C)&\simeq  \frac{\partial A}{\partial (v^{-p})}\Bigr\rvert_{v^{-p}=0} \ v^{-p} -2\kappa_-\sqrt{\frac{r_-}{r_--l}}e^{-\kappa_- v}\,.
    \end{split}
\end{equation}
Inserting these expressions in (\ref{eq117}), we obtain now,
\begin{equation}\label{eq129}
    \begin{split}
    |A(r;m_A)| &\simeq\left\rvert-1+\frac{1}{2\kappa_-}\sqrt{\frac{r_--l}{r_-}}\frac{\partial A}{\partial (v^{-p})}\Bigr\rvert_{v^{-p}=0} \ v^{-p}e^{\kappa_-v}\right\rvert|A(r;m_D)| \propto v^{-p}e^{\kappa_-v} 
    \end{split}\,,
\end{equation}
where we have considered the limit $v\rightarrow +\infty$.

Also in this case, the metric function $A(r; m_A)$ is unbounded at the Cauchy horizon. 
However, the derivation breaks down in the limit $l=r_-$, when the surface gravity goes to zero and the Cauchy horizon is located at the minimal length of the metric.

Here, a remark is in order. In our class of solutions for $F(R)$-gravity discussed in \S \ref{hh} the Cauchy horizon $r_-$ cannot coincide with the minimal length $l$ for positive values of $m$. However, our treatment is model-independent and the only requirement is that solution is in the form of (\ref{r1}) and admits two positive zeros for $A(r)$. In other words, any metric that corresponds to the Penrose diagram displayed in fig. \ref{penrose}.  For this reason, we conclude our analysis by investigating the mass inflation in the extremal limit $r_-=l$.  
%%%%%%%%%%%%%%%%%%%%%%

In this case by inserting the first order approximation of $A(r;m_B)$ at the Cauchy horizon,
\begin{equation*}
    A(r;m_B)\simeq \frac{d A}{d r}(r_-;m_B)(r-r_-)\,,
\end{equation*}
inside Eq. (\ref{eq119a}), we get
\begin{equation}\label{eq130}
    \frac{dr}{dv}\simeq \frac{1}{2}\sqrt{\frac{r-r_-}{r}}\frac{d A}{d r}(r_-;m_B)(r-r_-)\,,
\end{equation}
with the following solution,
\begin{equation}\label{eq131}
    \frac{2}{\sqrt{r-r_-}}\left(\sqrt{r}-\sqrt{r-r_-}\log\left[\sqrt{r-r_-}+\sqrt{r}  \right]  \right) \simeq \frac{1}{2}\frac{d A}{d r}(r_-;m_B)v\,.
\end{equation}
In the limit $r\rightarrow r_-$, this yields to an explicit expression for $(r-r_-)$ in terms of the advanced time, namely
\begin{equation}\label{eq132}
    r-r_-\simeq \frac{16r_-}{(d A/d r(r_-;m_B))^2}v^{-2}\,.
\end{equation}
Therefore, we can write $A(r;m_B)$ as
\begin{equation}\label{eq133}
    A(r;m_B)\simeq -\frac{16r_-}{| d A/dr(r_-;m_B)|}v^{-2}\,.
\end{equation}
Thus, aside the prefactors, while in Eqs. (\ref{AA1}) and (\ref{eq127}) the scalar function $A(r; m_B)$ decreases exponentially, Eq. (\ref{eq133}) predicts an asymptotic behavior proportional to $v^{-2}$.

Now we find,
\begin{equation}\label{eq134}
    A(r;m_C)\simeq \frac{\partial A}{\partial (v^{-p})}\Bigr\rvert_{v^{-p}=0} v^{-p} -\frac{16r_-}{|F_{,r}(r_-;m_B)|}v^{-2}\,,
\end{equation}
wich leads to the final result, 
\begin{equation}\label{eq135}
    |A(r;m_A)|\simeq\left\rvert-1+\frac{|d A/dr(r_-;m_B)|}{16r_-}\frac{\partial A}{\partial (v^{-p})}\Bigr\rvert_0 \ v^{-(p-2)}\right\rvert|A(r;m_D)|\,.
\end{equation}
In contrast to the previous discussions, this expression is bounded when $v\rightarrow+\infty$, since $p\geq12$ and the scalar function $A(r;m_D)$ remains finite due to the fact that $m_D\neq m_B$.
Therefore, we find that in the limit when the inner horizon coincides with the fundamental length-scale no exponential inflation of the metric function occurs.

\section{Conclusions}

In this paper we focused on a class of non singular (topological) black hole solutions in the framework of $F(R)$-gravity. These solutions are characterized by a null Ricci scalar and thanks to this fact are present in a wide class of $F(R)$-gravity models. We specifically required the presence of a minimal length in the metric, such that the radial coordinate is bounded and the space-time is free of singularity. The solution is presented in an analytical form and turns out to be the Reissner-Nordstrom space-time in the limit where the minimal length goes to zero. In general, we showed numerically that two horizons are always present and asymptotically we recover the Schwarzshild metric. These solutions can be obtained in vacuum and are not supported by exotic matter and may represent a valid alternative to the standard black hole representation.   
In this respect, we mention that the Kerr space-time still presents important conceptual problems and in the last years alternatives to the black holes have been subject to debate.

We investigated the motion of test particles and the problem of mass inflation. It is known that the presence of an inner horizon may lead to an exponential growth of the mass parameter of the metric. In GR the problem can be avoided in the case of regular black holes with a de Sitter core. In our case, we showed that if the minimal length of the metric coincide with the Cauchy horizon the mass inflation is not present and the solution is stable. Our findings are also supported by the very recent work of Rubio et al.\cite{Rubio:2021obb}.

A final comparison with other studies on mass inflation in Brans-Dicke theory of gravity may be useful to better understand the problem. In Ref. \cite{BD1} it is shown that in this case mass-inflation occurs in accreting black holes making small the variations of Brans-Dicke scalar inside the black hole itself. However, in Refs. \cite{BD2, BD3} it is shown that in Eddington-inspired Born-Infeld gravity, unlike in General Relativity, there is a minimum (critical) accretion rate below which there is no mass inflation.
\newline

\section*{Acknowledgements} We wish to thank A.\ Bonanno, S.\ Liberati, and L.\ Vanzo for useful discussions.

\end{document}